\documentclass{aastex}
\usepackage{amsmath,amssymb,graphicx}

\newcommand{\tmtextit}[1]{{\itshape{#1}}}

\shorttitle{Circularly polarized emission from the GCRT J1745$-$3009}
\shortauthors{Roy et al.}
\begin{document}

\title{Circularly polarized emission from the transient bursting radio source
GCRT J1745-3009 }
\author{Subhashis Roy}
\affil{NCRA-TIFR, Pune-411007, India}
\email{roy@ncra.tifr.res.in}

\author{Scott D. Hyman}
\affil{Department of Physics and Engineering, Sweet Briar College, Sweet Briar,
VA 24595}
\email{shyman@sbc.edu}

\author{Sabyasachi Pal}
\affil{International Centre for Radio Astronomy Research, University of
Western Australia, 35 Stirling Highway, Crawley, 6009, Australia}

\author{T. Joseph W. Lazio}
\affil{Remote Sensing Division, Naval Research Laboratory, Washington, DC
20375-5351}
\email{joseph.lazio@nrl.navy.mil}

\author{Paul S. Ray}
\affil{Space Science Division, Naval Research Laboratory, Washington, DC
20375-5352}
\email{Paul.Ray@nrl.navy.mil}

\author{Namir E. Kassim}
\affil{Remote Sensing Division, Naval Research Laboratory, Washington, DC
20375-5351}
\email{namir.kassim@nrl.navy.mil}

\begin{abstract}

We report detection of strong circularly polarized emission from the transient
bursting source GCRT J1745-3009 based on new analysis of 325 MHz GMRT
observations conducted on 28 September 2003. 
We place 8 R$_{\odot}$ as the upper limit on the size of the emission region. 
The implied high brightness temperature required for an object beyond
1 pc and the high fraction of circular polarization firmly establish the
emission as coherent. Electron cyclotron or plasma emission from a highly
subsolar magnetically dominated dwarf located $\le$4 kpc away could have given
rise to the GCRT radio emission.
\end{abstract}


\section{Introduction:}

The present generation of largely parabolic dish-based radio interferometers
and data analysis procedures are poorly optimized for discovering transient
emission. Consequently, the characteristics of transient radio source
populations remain highly undersampled. Nevertheless, dedicated programs
are revealing novel radio transients with unique emission characteristics,
including GCRT J1745$-$3009. Discovered in 325 MHz archival VLA data taken in
September 2002, this source exhibited $\sim$10 minute long, 1 Jy peaked bursts
with a $\sim$77 minute period over the $\sim$7 hour baseline available in the
discovery data \citep{HYMAN2005}. 
The measured time scales of the bursts implied a brightness temperature
exceeding the 10$^{12}$ K Compton limit for a distance beyond $\sim$100 pc. As
discussed in \citet{HYMAN2005}, its characteristics did not match any known
mechanisms of emission in transient compact sources.
As a result, GCRT J1745$-$3009 appeared to represent a member of a new class of
coherently emitting objects.

Search to detect additional bursts from the source in archival data sets
were made, resulting in its re-detection for $\sim$2 minutes at 
325 MHz Giant metrewave radio telescope (GMRT) observation
in September 2003 \citep{HYMAN2006} and in March 2004 \citep{HYMAN2007}.  
At the last known epoch of emission detected in 2004, the source exhibited an
unusually steep spectrum with alpha = $-$13$\pm$3 (S ($\nu$)$\propto
\nu^{\alpha}$) \citep{HYMAN2007}.

Several theories have been proposed to explain the emission from GCRT
J1745-3009 (see \citet{HYMAN2007} and references therein). These include
nulling \citep{KULKARNI2005}, double \citep{TUROLLA2005}, precessing
\citep{ZHU2006} and a transient white dwarf pulsar \citep{ZHANG2005}. Also,
\citet{HALLINAN2007} suggested a nearby ultracool dwarf as its
progenitor. Unfortunately, non-detection of GCRT J1745$-$3009 at
frequencies other than 325 MHz \citep{KAPLAN2008} hindered attempts to discover
its progenitor.  Its brightness temperature in possible excess of the Compton
limit suggests coherent processes such as electron cyclotron maser emission or
plasma emission. Notably, both of these mechanisms produce circularly
polarized emission, motivating us to re-examine earlier detections of GCRT
J1745$-$3009 in search of its signature. We have reanalyzed the 325 MHz GMRT
data from September 2003, the detection with the highest signal to noise ratio
per integration time available.  The subsequent sections are arranged as
follows - in Sect. 2 and 3 we describe our reanalysis procedure and its
results, respectively. The interpretation of these results are presented in
Sect. 4 and conclusions in Sect. 5.

\section{Observations and data analysis:}

During the observations, the pointing center of the GMRT antennas was
0.5$^{\circ}$ West of GCRT J1745$-$3009 (hereafter GCRT). The unpolarized
calibrator 3C48 was observed 
for absolute gain calibration - see Hyman et al. 2006 for further
details of the observation and total intensity calibration. At the time of
these observations, the full polarization mode of the GMRT was unavailable
and only parallel-hand correlations (Stokes RR and LL) were recorded.
If the polarization leakage coefficients of antenna `i' is denoted by D$_i$,
then the observed right circularly polarized signal on the $i$-$j$ baseline is
\begin{equation}
	R'_i.R'^{\ast}_{j} = (1 - D_i - D_j^{\ast}).R_i.R^{\ast}_j + R_i.(D_j
L_j)^{\ast} + (D_i.L_i).R_j^{\ast} + (D_i R_i).(D_j.R_j)^{\ast}+ (D_i.L_i).
(D_j.L_j)^{\ast}
\end{equation}
In general, the quantities are complex; we denote the `true' values without
primes and the measured quantities with primes; and a similar equation holds
for the left circularly polarized signal $L^\prime_i{L^\prime_j}^*$.
During standard calibration without full polarization information, the
leakage terms are considered small and discarded. However, the above
equation shows that even for an unpolarized source the last two terms [$(D_i
R_i).(D_j R_j)^{\ast} +(D_i L_i) . (D_j L_j)^{\ast}$] lead to amplitude closure
errors.  If the source is linearly polarized ($R_i . L_j^{\ast} \neq 0$),
additional errors are added. 

To determine magnitudes of these leakage coefficients, the connections for the
2 polarizations of a single antenna is reversed. This allows acquiring cross
products of opposite handed polarizations of one antenna with the rest of the
antennas. Observations of strong unpolarized calibrators have shown
that $D_i \lesssim 0.1$ at~325~MHz.  Since
the leakage terms for different antennas are expected to be uncorrelated,
the estimated error in Stokes V in the image plane for even a 100\% linearly
polarized source will be $<$ 5\% of Stokes I.  Moreover, the polarization
leakage in antennas cannot add coherently in this case, indicating absence of
any significant spurious small diameter source in Stokes V in the image plane.
However, the above analysis is valid for a source along the axis of the
antennas and the GCRT was off-axis by 0.5$^{\circ}$.
Antenna off-axis positions may suffer from correlated patterns among the
antennas (e.g., different FWHM of the right and left circularly polarized
primary beams) leading to non Zero Stokes V. Therefore,
we obtained supplementary polarization calibration observations of the
unpolarized calibrator 3C48 at 325 MHz with 20 different angular offsets
within the FWHM of the antenna primary beam. 
Fig. 1 shows that the pointing positions sampled a range of primary beam offsets
extending beyond the GCRT offset as observed on 2003 Sep 28.  These
observations were sufficiently short such that rotation in feed parallactic
angle within any particular pointing position were negligible. For all the
off-axis positions after calibration, the maximum value of Stokes V was found
to be 0.02 times Stokes I. Hence, the error in Stokes V within the
FWHM of the antennas is $\leq$2\% of Stokes I for an unpolarized source and
$<$5\% for a high linearly polarized source at 325 MHz band of GMRT.

\begin{figure}[h]
  \resizebox{!}{6cm}
  {\includegraphics{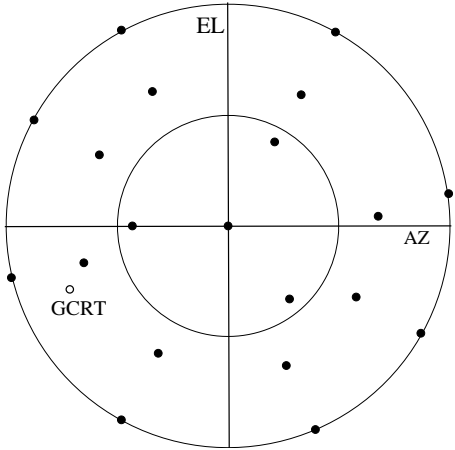}}
  \caption{3C48 observed through 20 different parts of the primary
  beam. The outer circle indicates the FWHM of the primary beam (radius 45$'$)
  and the inner circle is half of that distance. Azimuth is along X-axis, and 
  elevation along Y-axis.The location of the GCRT is indicated by an open
  circle.}
\end{figure}

A variant of multi-resolution Clean was used to better image the non-variable
and extended field sources.  
After amplitude and phase self-calibration of non-variable sources, all sources
except one were subtracted from the \tmtextit{uv}-data and the inverse of the
amplitude self-calibration table applied to avoid changes in the flux density
of the GCRT during imaging. Subsequently, the GCRT
and the strong compact field source (G358.917+0.073, located 0.2$^{\circ}$ away
from the phase center) not subtracted in the above step were imaged
simultaneously from every time sample in the data to confirm the amplitude
stability of non-variable sources.
The rms noise in the maps are $\sim$34 mJy.beam$^{- 1}$ in each polarization.
The equivalent Stokes I noise $\sim$25 mJy.beam$^{- 1}$ is approximately a
factor of 2 lower than in Hyman et al. 2006. We attribute the significant
reduction in noise due to better removal of extended sources before imaging GCRT
coupled with more aggressive inner \tmtextit{uv} cutoffs (3.3 k-Lambda vs. 2
k-Lambda employed by Hyman et al. 2006).

\section{Results}
Light curves of the unresolved source, the GCRT, during its active state on September
28, 2003 along with the unrelated compact field source G358.917+0.073 using
their peak flux densities are shown in Fig.  2 (top panel) for both the
polarizations. 
\begin{figure}[h]
  \resizebox{!}{12cm}{\includegraphics{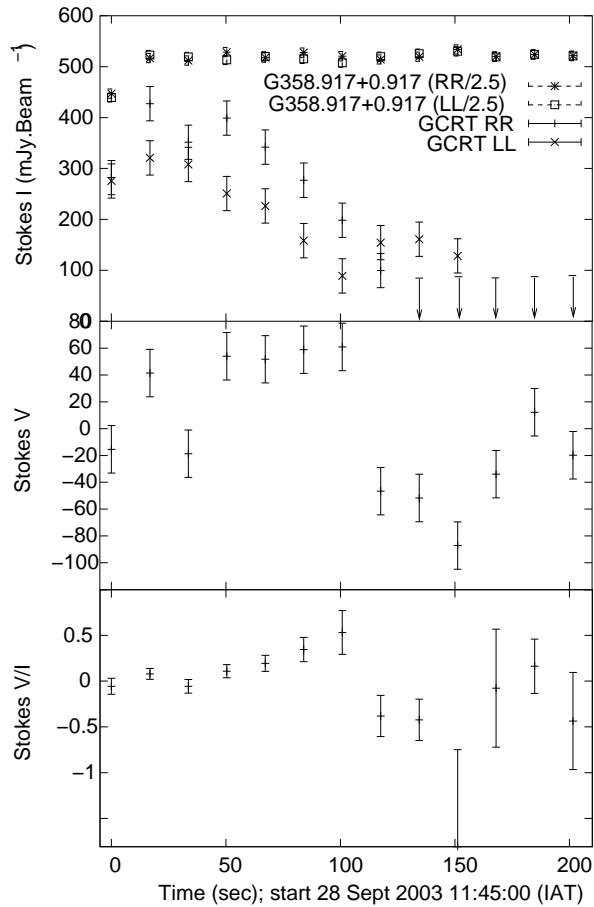}}
  \caption{Light curves of the GCRT and an unrelated compact field source
  G358.917+0.073 scaled down by 2.5 for both RR and LL (top). The arrows are
  3$\sigma$ upper limits for non-detections. Other 2 panels show Stokes V at
  the same timestamps as above (middle) and the ratio of Stokes V to I
  (bottom).}
\end{figure} 

Due to a correction in data timestamps produced by upgraded observatory
software, the time used in this paper lags by 7.4 sec relative to
\citet{HYMAN2006}. 
The peak flux density of both the sources are smaller for the first data point,
which is erroneously reduced by the data acquisition software at the start of
data acquisition scans. 
The plot shows that while the flux density for both the
polarizations of the unrelated field source are equal and steady within the
error-bars, the flux density of the GCRT varies systematically with time. At
the start of the scan, the RR flux density is higher than the LL, but it drops almost
exponentially to zero after about 200 seconds and after which the GCRT could
not be identified.
Due to lower rms noise than \citet{HYMAN2006}, we have been able to detect the
source for $\sim$1.5 times longer duration than reported earlier.

\citet{HYMAN2006} searched for circular polarization from this source in a map
made using data averaged over the time scale of emission and derived an upper
limit of 15\%.  However, because the difference between the two polarizations
changes sign within the averaging time, their ability to constrain stokes V
was smeared out. Adopting their approach during our re-analysis yields a stokes
V of 12\%, consistent with their reported limit.  More recent reanalysis of its
2002 outburst \citep{SPREEUW2009} yields an upper limit of 8\% on the fraction
of circular polarization averaged over the emission timescale.
To study the variation of the circular polarization, we have also made maps of
the GCRT in Stokes V, and a light curve made from these maps is shown in Fig. 2
(middle panel). In Stokes V, the rms noise in the map is $\sim$17
mJy.beam$^{-1}$, quite close to the expected thermal noise.\\
The figure shows that the sign of Stokes V reverses after about 100 sec from
the start of the scan and within the data averaging time of
$\sim$17 sec. If the fastest variation in the source properties approach the
velocity of light, the emission region would be smaller than
8~R$_{\odot}$. Fig. 2 (bottom panel) shows the ratio of Stokes V to I obtained
by dividing the Stokes V maps by the Stokes I maps of the same timestamps using
COMB in Aips. It shows the fractional circular polarization is initially low,
but increases with time. After 100 seconds from the start of the scan it
reverses sign. The fraction of circular polarization approaches $\sim$100\%
about 150 seconds from the start of the scan.  Fig. 3 shows a gray scale map
of the difference of Stokes V images made from data integrated between
35 to 103 and 103 to 155 seconds from the start of scan, respectively, during
which the sign of stoves V reversed. This result ($>$6 $\sigma$) represents our
highest significance detection of Stokes-V from the GCRT.
\begin{figure}[h]
\begin{minipage}{0.48\textwidth}
\includegraphics[width=0.96\textwidth,clip=true,angle=0]{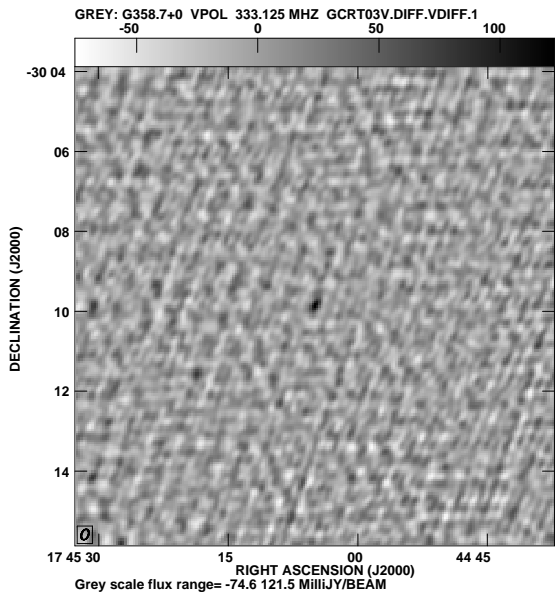}
\caption{Stokes V difference image made from data integrated between
35 to 103 and 103 to 155 seconds respectively from the start of scan and
bridging the polarization reversal. Rms noise 19.2 mJy.Beam$^{-1}$. The
location of the GCRT is at the center.} \end{minipage}
\end{figure}

Due to large closure errors on the calibrators with lower side band data from
the correlator, the bandwidth of the present analysis was limited to 16 MHz.
The measured Stokes I spectral index ($S \propto \nu ^{\alpha}$) 
is $-4\pm 3$, which is better constrained than in \citet{HYMAN2006} ($-4\pm
5$).
We have also reanalyzed the September 2002 and March 2004 GCRT detections.
However, we could not detect any significant circular polarization from those 2
epochs, and our upper limits
remain as in \citet{HYMAN2005} and \citet{HYMAN2007} respectively. 

\section{Discussions}

\subsection{Emission mechanism}

From the measured flux density and the decay time scale ($\sim 2$ min) of the
2002 bursts, \citet{HYMAN2005} placed an upper limit of 70 pc, which has been
revised to 14 pc \citep{SPREEUW2009} for incoherent emission.  Adopting the
$\sim 17$ sec time scale for polarization reversal derived from our new
analysis of the detected 2003 burst as the fastest variation yields an upper
limit of 8 R$_{\odot}$ on its size. 
Assuming the above size of emission region in 2002 emission lowers the distance
limit to $\sim 10$ pc.  Given a very low probability of encountering an object
within 1 pc and taking this as a distance lower limit still implies a
brightness temperature $>$10$^{10}$ K.  Such a high brightness temperature and
circular polarization fraction reaching $\sim$100\% rules out thermal emission
and incoherent non-thermal emission like synchrotron and gyro-synchrotron
\citep{MELROSE1982}, and indicates a coherent emission mechanism.  Coherent
radiation is known to arise from 3 different processes, (i) pulsar emission,
(ii) electron cyclotron maser emission and (iii) plasma emission, which are
discussed below.

Pulsar emission could produce highly circularly polarized emission within a
single pulse. However, the integration time (16.8 sec) used in the data is
longer than the pulse period of any known pulsar, and circular polarization
fraction of pulsars goes down when averaged over a full pulse and is typically
found to be significantly less than 100\% near an observing frequency of 325
MHz (D. Mitra, private communication). This argues against any pulsar-based
model as an explanation for the observed properties of the GCRT.\\ 
Electron cyclotron maser emission occurs at the electron gyro-frequency and its
harmonics and could also produce highly circularly polarized emission
\citep{DULK1985}.  If the GCRT emission at 330 MHz was due to cyclotron
emission at its fundamental frequency, the required magnetic field is $\sim$120
Gauss and $\sim$2 lower at the second harmonic. These are typical values of
magnetic field in a stellar corona. 
The relative bandwidth of emission is $\Delta \nu / \nu \approx v_0^2 /
c^2$ for electrons with speed $v_0$, and is typically $\sim 0.01 - 0.1$
\citep{VAN-DEN-OORD1994}. We note, however, that the individual spikes of
emission could have durations much smaller than our integration time and 
could have occured from different physical locations.  A significant
variation of magnetic fields within these spikes of emission would cause
a variation of emitting frequencies causing a broadening of observed bandwidth.
This could account for $\gtrsim$30 MHz bandwidth observed in \citet{HYMAN2007}.
The very steep spectral index reported in \citet{HYMAN2007} could also be
explained if the peak of emission at that time occured at $\lesssim$310 MHz. \\ 

Plasma emission occurs at the plasma frequency or one of its lower order
harmonics. It involves plasma instabilities and subsequent conversion of a
fraction of its energy to electromagnetic radiation. Different mechanisms of
generating plasma instabilities are known \citep{VAN-DEN-OORD1994}.  Common
characteristics of these are narrow band emission with a high level
(approaching 100\%) of circular polarization and high brightness temperature at
meter wavelengths. The typical bandwidth of plasma emission is $\Delta \nu / \nu
\sim 0.01$. However, as described above, there could be several spots of
emission slightly differing in frequency, which would also explain a bandwidth
of $>$32 MHz as reported in \citet{HYMAN2007}. We note that plasma frequency of
330 MHz corresponds to an ion density of $\sim$10$^9$ ions.cm$^{-3}$, which is
less likely but cannot be ruled out in the corona of a brown dwarf.
However, untill the magnetic field strengths or ion density could be measured
near the object, any electron cyclotron or plasma emission mechanism
attributable to the object cannot as yet be firmly established.

\subsection{Distance to the GCRT and its classification}

If the upper limit to the brightness temperature ($T_B$) and the linear size
($L$) of the emission region is known, an upper limit to its distance (D) can
be calculated.  For electron cyclotron maser emission, it is found that $T_B$
could reach 10$^{20}$ K as in auroral kilometric radiation \citep{ERGUN2000}.
This upper limit has also been estimated theoretically \citep{MELNIK1994}. For
plasma emission, $T_B$ is expected to be less than 10$^{16}$ K
\citep{STEPANOV1999, STEPANOV2001}.
However, for a certain type of plasma emission from double layers, $T_B$ could
reach 10$^{25}$ K \citep{KUIJPERS1989}. Therefore, we could easily consider the
upper limit on its $T_B$ to be 10$^{20}$ K. With L $<$8 R$_{\odot}$
(Sect. 4), the upper limit on its distance is $\sim 100$ kpc, which indicates
that the emission does not originate from another galaxy. However, this limit
can be much improved by considering known sources of circular polarisation in
the Galaxy, which are not ruled out already. 

Most of the known low frequency emitting objects are either synchrotron or
thermal sources, from which the circular polarization fraction is less than a
few percent.  However, flare stars in the Galaxy could emit strong high
circularly polarized emission. These objects are typically dwarfs stars of
class G to M (mostly M).
Some of them could be of size $\sim$0.15 R$_{\odot}$ 
(e.g., UV Cet with spectral type M6). As described in \citet{GERSHBERG2005},
more than 50 observations have been carried out in radio bands towards
them from the early days of radio astronomy in 1958 to 2001 and cover a
frequency range from 20 MHz to several GHz. 
The flux densities recorded for these flares at cm wavelengths
are typically in millijanskys, and the highest recorded was only a fraction of
a Jy. However, at lower frequencies, particularly at decameter wavelengths,
flux densities of some of the flares recorded are $\sim$100 Jy
\citep{ABDUL-AZIZ1995} to $\sim 10^4$ Jy \citep{SLEE1963}.  The highest
luminosity of any flare star(s) in the metre wave band is $\sim$230 Jy at 136
MHz as recorded by \citet{SLEE1969} towards the Orion nebula located
$\sim$400 pc away.  If the luminosity of the GCRT in the 2002 outburst is the
same, it would be located at a distance of $\sim$4 kpc. 

Earlier observations of flare stars showed wide variation in the circular
polarization fraction with some of the bursts being unpolarized while others
having circular polarization fraction reaching up to 100\% (see e.g.,
\citet{ABADA-SIMON1994}). At meter wavelengths, \citet{NELSON1979} observed 40
to 60\% circularly polarized emission in 8 cases from flare stars.
\citet{LANG1983} reported observations of AD Leo with circular polarization
fraction of about 15\%, and the emission properties were explained
as electron cyclotron emission.  Reversal in the sense of circular
polarization was also reported from the above flare star \citep{JACKSON1989}.
As discussed in \citet{DULK1985} and \citet{MELROSE1980}, mode coupling could
prevent attainment of a high polarization, which could operate for the GCRT
when the circular polarization fraction is small and also explain our observed
polarization reversal in the data described above.

Measured Rotation period for stars show a decrease in rotation period towards
lower mass (0.1 M$_{\odot}$) objects  reaching $\sim$0.1 day (e.g. in
Pleaides) \citep{IRWIN2008}. A few of these ultracool dwarfs
produce flares at radio frequencies \citep{BURGASSER2005}.
As in \citet{HALLINAN2007}, within one rotation of the object, there could be
two pulses of emission. If this is the case during the GCRT 2002 outbursts, its
rotation period is $\sim$0.1 day, consistent with the above finding. Hence,
an ultracool dwarf could be progenitor of the GCRT.
\citet{KAPLAN2008} have considered the object `C' as a possible progenitor of
the GCRT. However, emission from an ultracool dwarf (L4.5V) do not produce a
good fit to the observed multi-band photometry of the object C.  We find the
goodness of the fit 
do not change significantly (rms error $\sim$0.3 magnitude) from
\citet{KAPLAN2008} results when the suggested K7V star is replaced by a
$\sim$0.1 M$_{\odot}$ star $\sim$0.1 Gyr old \citep{BARAFFE1998}.  This could
be a young mid to late M type of star with A$_V$ $\sim$3.5, indicating it to be
at a distance of $\sim$4 kpc.  Such a distance matches the 10$^{20}$K cyclotron
brightness temperature limit (radius $\sim$0.15 R$_{\odot}$). However,
considering the uncertainty ($\sim$0.1 magnitude) in the IR magnitudes
\citep{KAPLAN2008} and in the theoretical model, the quality of the fit is
marginal and object `C' is considered only as a candidate progenitor
of the GCRT.  We, however, note that a hithertho undetected ultracool dwarf
located $\gtrsim$200 pc \citep{KAPLAN2008} could also be the progenitor of the
GCRT.

\section{Conclusions}

We have detected time-varying, highly circularly polarised emission from the
GCRT J1745-3009 based on a re-analysis of its outburst on 28th September 2003.
The percentage of circular polarization approaching $\sim$100\% commensurate
with high brightness temperature ($>$10$^{10}$K for a source $>$1 pc away)
significantly strengthens the original suggestion of a coherent emission
scenario \citep{HYMAN2005}, but its properties are inconsistent with a pulsar
origin.
The time scale of change in sign of the circular polarization allows us to
estimate an upper limit of 8 R$_\odot$ on its size.  Based on its measured
emission characteristics, either electron cyclotron maser or plasma emission
processes could be the most plausible origin scenarios.  

Ascribing the 77 minute periodicity observed in its 2002 outbursts to half the
rotation period of its progenitor, observed radio emission of the GCRT could
be explained as outbursts from a highly subsolar flare star. Its distance is
estimated to be $\le$4 kpc. Independent verification, would however, be
required to confirm its progenitor.

\section*{Acknowledgments}
We thank the staff of the GMRT that made these observations possible. GMRT is
run by the National Center for Radio Astrophysics of the Tata Institute of
Fundamental Research. S.D.H. is supported by funding from Research Corporation.
Basic research in radio astronomy at the Naval Research Laboratory is supported
by 6.1 base funding.

\bibliographystyle{aa}

\end{document}